\documentclass{aa}
\usepackage{graphicx} 
\usepackage{txfonts} 
\usepackage{natbib}
\bibpunct{(}{)}{;}{a}{}{,} 
\usepackage[english]{babel} 
\usepackage{url} 
\usepackage{color} %
\usepackage{amsmath}
\usepackage{bm}
\usepackage[normalem]{ulem}
\usepackage[toc,page]{appendix}
\usepackage[version=3]{mhchem}
\usepackage{color}
\usepackage{colortbl}
\usepackage{mathtools}
\usepackage{xcolor}
\usepackage{chemfig}
\titlerunning{NH$_{3}$ snow line and binding energy} 
\authorrunning{S Kakkenpara Suresh et al}
\graphicspath{{./}{Figures/}}
\usepackage{float} 
\usepackage{adjustbox} 
\usepackage{blindtext}
\usepackage{comment}
\usepackage[labelfont=bf]{caption} 
\usepackage{subcaption}
\usepackage{enumitem}
\usepackage{hyperref} 
\hypersetup{
    colorlinks=true,
    linkcolor=blue,
    filecolor=blue, 
    citecolor=blue, 
    urlcolor=blue, 
    pdftitle={Overleaf Example},
    pdfpagemode=FullScreen,
    } 
\usepackage[toc]{appendix}
\usepackage{gensymb} 
\usepackage{booktabs}

\begin{document} 
\pagenumbering{arabic}

   \title{Experimental study of the binding energy of NH$_{3}$ on different types of ice and its impact on the snow line of NH$_{3}$ and H$_{2}$O
 }

   \subtitle{}

   \author{S. Kakkenpara Suresh
          \inst{1,}\inst{2}
          \and
          F. Dulieu\inst{1}
          \and 
          J. Vitorino\inst{1}
          \and
          P. Caselli\inst{2}
          }

   \institute{LERMA, CY Cergy Paris University,
              5 mail Gay Lussac, Neuville-sur-Oise\\
              \email{shreyaks@mpe.mpg.de}
         \and
             Max Planck Institute for Extraterrestrial Physics, Giessenbachstraße 1, 85748 Garching, Germany \\
             }

   \date{Received; accepted}

 
  \abstract
   {N-bearing molecules (like N$_{2}$H$^{+}$ or NH$_{3}$) are excellent tracers of high-density and low-temperature regions, such as dense cloud cores.  Moreover, they could shed light in understanding snow lines in protoplanetary disks and the chemical evolution of comets. However, a lot remains unknown about the chemistry of these N-bearing molecules on grain surfaces - which could play an important role in their formation and evolution.  }
   {In this work, we experimentally study the behaviour of NH$_{3}$ on surfaces that mimic grain surfaces under interstellar conditions in the presence of some of the other major components of interstellar ices (i.e. H$_2$O, CO$_2$, CO). We measure the binding energy distributions of NH$_{3}$ from different H$_{2}$O ice substrates and also investigate how it could affect the NH$_{3}$ snow line in protoplanetary disks.}
   {We performed laboratory experiments using the Ultra High Vacuum (UHV) setup VENUS (VErs des NoUvelles Syntheses) where we co-deposited NH$_{3}$ along with other adsorbates (here H$_{2}$O, $^{13}$CO and CO$_{2}$) and performed Temperature Programmed Desorption (TPD) and Temperature Programmed-During Exposure Desorption (TP-DED) experiments. The experiments were monitored using a Quadrupole Mass Spectrometer (QMS) and a Fourier Transform Reflection Absorption Infrared Spectrometer (FT-RAIRS). We obtained the binding energy distribution of NH$_{3}$ on Crystalline Ice (CI) and compact-Amorphous Solid Water ice (c-ASW) by analysing the TPD profiles of NH$_{3}$ obtained after consequent depositions on these substrates. }
   {In the co-deposition experiments, we observe a significant delay in the desorption and a decrease of the desorption rate of NH$_{3}$ when H$_{2}$O is introduced into the co-deposited mixture of NH$_{3}$-$^{13}$CO or NH$_{3}$-CO$_{2}$, which isn't the case in the absence of H$_{2}$O. Secondly, we notice that H$_{2}$O traps roughly 5-9$\%$ of the co-deposited NH$_{3}$, which is released during the phase change of water from amorphous to crystalline. Thirdly, we obtain a  distribution of binding energy values of NH$_{3}$ on both the ice substrates instead of an individual value as assumed in previous works. For CI, we obtained an energy distribution between 3780K - 4080K, and in the case of amorphous ice, the binding energy values are distributed between 3630K - 5280K - in both cases using a pre-exponential factor of A = 1.94$\times$10$^{15}$s$^{-1}$.}
   {From our experiments, we conclude that the behaviour of NH$_{3}$ is significantly influenced by the presence of water owing to the formation of hydrogen bonds with water, in line with quantum calculations. This interaction, in turn, preserves NH$_{3}$ on the grain surfaces longer and to higher temperatures making it available closer to the central protostar in protoplanetary disks than previously thought. It well explains why NH$_{3}$ freeze out in pre-stellar cores is efficient. When present along with H$_{2}$O, CO$_2$ also appears to impact the behaviour of NH$_{3}$ retaining it at temperatures similar to those of water. This may impact the overall composition of comets, particularly the desorption of molecules from their surface as they approach the Sun.}


   \maketitle
%

\section{Introduction}
Ammonia is one of the six major molecules found in the solid phase in interstellar ices \citep{Boogert2015}. It has been observed in a variety of environments in the universe like comets (e.g. \cite{poch2020ammonium}), star-forming regions (e.g., \cite{feher2022ammonia}), external galaxies (e.g. \cite{gorski2018survey}), the centre of our galaxy (e.g. \cite{sandqvist2017odin}) and in the Solar System planets. It was first detected in the interstellar medium towards the Galactic centre by \cite{PhysRevLett.21.1701} by its J=1, K=1 inversion transition. In dense starless clouds, where temperatures can be as low as 6K \citep{crapsi2007observing, 2007A&A...467..179P} and number densities between 10$^{4}$ - 10$^{6}$ cm$^{-3}$ \citep{keto2010dynamics}, molecular gas tracers like CO and CS are depleted from the gas phase and mainly reside on the surface of dust grains upon freeze-out \citep{caselli1999co, tafalla2002systematic}. A similar scenario (large degree of C-bearing molecular freeze-out) is expected in the mid-plane of protoplanetary disks, where number densities are orders of magnitude larger than the central regions of dense starless cores \citep{dutrey1997chemistry,henning2013chemistry, qi2013imaging}.  However, N-bearing molecules, in particular, NH$_{3}$ and N$_{2}$H$^{+}$ and their deuterated forms, appear to be more resilient to freeze-out (e.g. \cite{caselli2002dense}, \cite{tafalla2002systematic}, \cite{tafalla2004internal}, \cite{crapsi2007observing}). For this reason, they are considered excellent tracers of dense and cold interstellar regions. More recent work, using multi-transition studies done with the IRAM 30m antenna and high sensitivity interferometric observations with ALMA and JVLA, has shown that these molecules freeze out within the central region of pre-stellar cores (e.g. \cite{redaelli2019high}, \cite{caselli2022central}, \cite{pineda2022interferometric}), although at higher densities than CO. Gas-grain astrochemical models are now able to reproduce the observations \citep{caselli2022central, pineda2022interferometric}. Still, they are limited by uncertainties in certain factors like the binding energies of NH$_{3}$ and knowledge of its surface chemistry. Following the work of \cite{collings2004laboratory}, \cite{penteado2017sensitivity} derived binding energies for several species, including NH$_{3}$. Similarly, \cite{collings2004laboratory}, \cite{he2016binding} and \cite{suhasaria2015thermal} have studied the desorption properties of NH$_{3}$ from various substrates. However, the binding energies of NH$_{3}$, the importance of surface chemistry for its formation, and its chemical desorption efficiency \citep{caselli2017nh3, sipila2019does} can be better constrained.


 Ammonia has been shown to be stored efficiently in the form of ammonium salts by \cite{kruczkiewicz2021ammonia}, which at higher temperatures decompose to release ammonia into the gas phase irrespective of the presence or absence of water. This release into the gas phase occurs at a temperature higher than that of the desorption of water ice (Temp $\sim$ 154K) but lower than room temperature. This implies that ammonia could be found closer to a young stellar object than the water snow line. The majority of NH$_{3}$ in molecular mantles should therefore be free to interact with the major components of interstellar or cometary ices such as water, CO$_2$ or CO. Similarly, \cite{poch2020ammonium} have demonstrated that ammonium salts are a dominant nitrogen-reservoir on cometary surfaces, explaining the low measured cometary nitrogen-to-carbon ratio as compared to that of the Sun. Finally, as ammonia ice is one of the major nitrogen reservoirs in star-forming regions \citep{2011ApJ...740..109O}, knowledge of the ammonia snow line is crucial while considering the formation and evolution of more complex N-bearing species in planetary systems.

The aim of this work is to study the behaviour of ammonia in the presence of the different components of molecular ice mantles like H$_{2}$O, CO and CO$_{2}$. We, then, performed studies to obtain the binding energy of ammonia on different water ice substrates, namely c-ASW and CI. Additionally, we conducted experiments to determine the snow line of ammonia. The article is organised as follows: Section 2 describes the experimental set-up; section 3 details the experiments and the results; section 4 discusses the conclusion and section 5 outlines the discussion and future prospects. 

\section{Experimental Set-up}

The experiments are performed on the VENUS (VErs de NoUvelles Syntheses) setup, allowing us to reproduce the conditions in cold, dark clouds \citep{Congiu2020}. VENUS consists of an ultra-high vacuum (UHV) chamber - that can attain a base pressure of 1$\times$10$^{-10}$ hPa (1 hPa = 1 mbar) - and five independent beams (which can be used simultaneously to inject the desired species). A Quadrupole Mass Spectrometer (QMS) and the beam of a Fourier-Transform Reflection Absorption Infrared Spectrometer (FT-RAIRS) are also present in the UHV chamber (hereafter referred to as the main chamber) that can be used to make qualitative and quantitative measurements during experiments. Depositions are made on a polycrystalline gold substrate - enclosed within the main chamber - which is chemically inert and acts as a proxy surface to dust grains for species adsorption/desorption for reactions. This surface is attached to the cold head of a closed-cycle He cryostat, which allows the surface temperature to be varied between 10 - 400K. Four of the five beams are separated from the main chamber via two intermediary chambers - 1 and 2. These intermediary chambers are responsible for regulating the pressure of the injected species between the four beams (\textit{p} $\sim$ 10$^{-5}$-10$^{-4}$ mbar) and the main chamber via differential pumping. The four beams are labelled: the top beam, the central beam, the right beam, and the bottom beam. For the experiments in this paper, only the first three beams are used. The fifth beam is attached to the main chamber and can inject species directly into the chamber (also known as \textit{background deposition}).

\begin{figure*}[h!] 
\centering
\includegraphics[width=16cm]{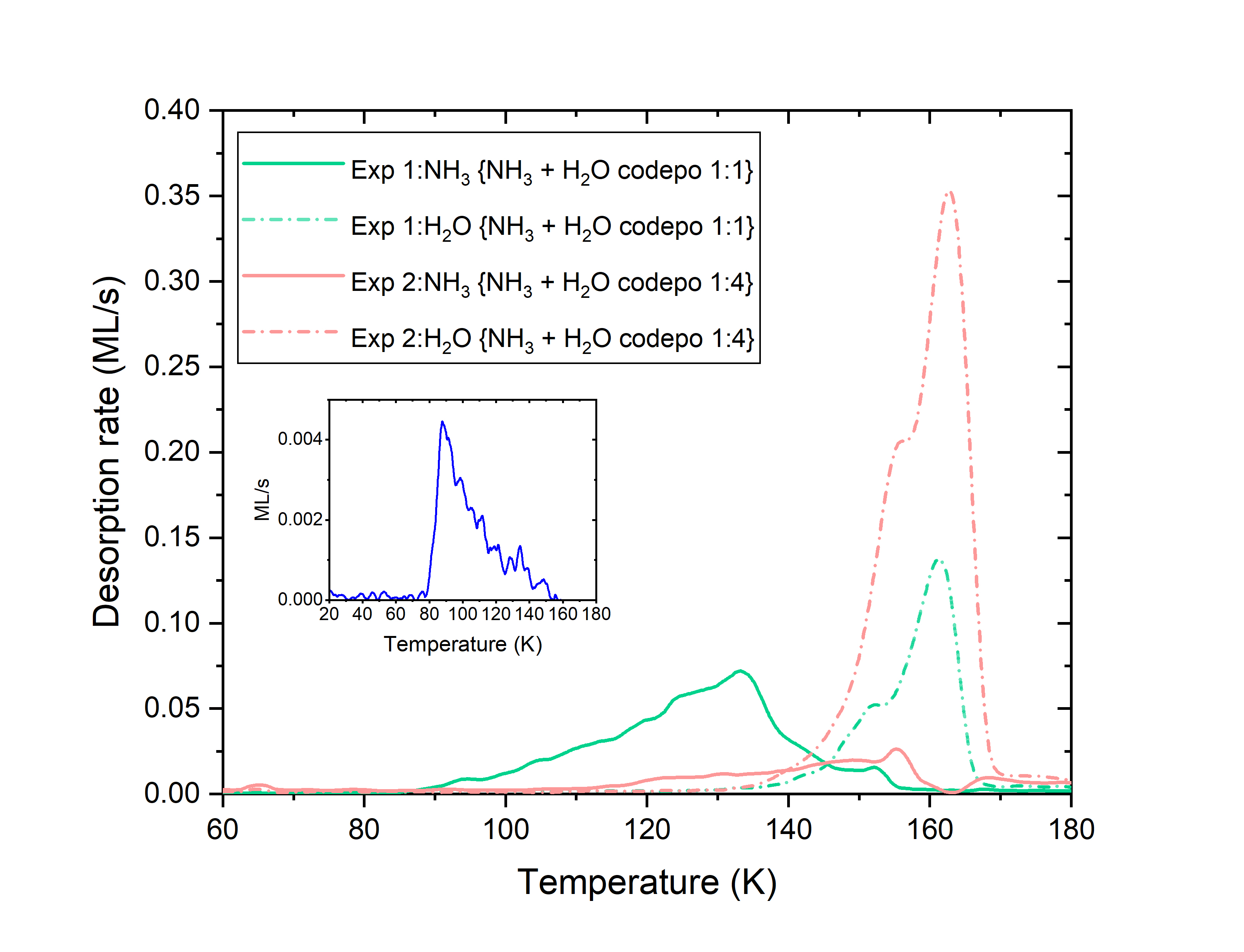} 
\vspace*{-8mm} 
\caption{\small NH$_{3}$-H$_{2}$O co-deposition experiments. All experiments have been performed on a gold substrate. The TPDs have a ramp of 0.2K/s. The solid lines represent the desorption of NH$_{3}$ while the dashed-dotted lines represent the desorption of water. Lines of the same colour belong to the same set of experiments. Inset: TPD of NH$_{3}$ from a gold surface used to calibrate all the consequent experiments.}
\label{fig:fig1}%
\end{figure*}

\section{Results}
\subsection{Co-deposition experiments}

The initial set of experiments aims to understand the interaction of ammonia with the major components found on the grain mantles, as mentioned earlier. Our work considers three major molecules: H$_{2}$O, CO, and CO$_{2}$. However, for our experiments, we use $^{13}$CO (mass = 29 amu) instead of $^{12}$CO (mass = 28 amu) to allow clear distinction from atmospheric N$_{2}$ (mass = 28 amu) by the QMS. Similarly, the major mass channels for water is 18 amu and that of NH$_{3}$ is 17 amu. However, water also has fragments of mass = 17 amu and NH$_{3}$ fragments into mass = 16 amu. To distinguish between these overlapping fragments of H$_{2}$O and NH$_{3}$, we independently perform experiments using each species, determine the ratio/percentage of fragmentation into each mass by the QMS and calibrate the quantities for each mass proportionately. Initially, we co-deposit ammonia with each of these components separately. We inject the species onto the gold substrate, maintained at a temperature of 10K during the deposition. Each of these species is dosed using separate beams. A laser beam is used to ensure that every beamline overlaps on the same spot on the deposition substrate. The details of the alignment process can be found in \cite{Congiu2020} and have been omitted here for brevity. The pressure of the NH$_{3}$, $^{13}$CO and CO$_{2}$ at first stage of the beam is a few 10$^{-5}$mbars (corresponding to a flux: 0.1 sccm \footnote{sccm = standard cubic centimetre}) while that of H$_{2}$O is roughly 4.8x10$^{-5}$ mbar. We define a monolayer as 1$\times$ 10$^{15}$ molecules cm$^{-2}$. The reproducibility between two depositions is within a few per cent and the accuracy in the absolute determination of an ML is around 20$\%$ since it is more precise for some molecules and less for other species (like NH$_{3}$). Table  \ref{table:Table1} lists the experiments used for the co-deposition experiments along with the ratios and monolayers of each species used. The raw data (mass spectra as well as IR spectra) are accessible online via a dedicated database\footnote{\url{https://lerma.labo.cyu.fr/DR/traitement.php}}. We have also carried out and analysed other sets of experiments under similar conditions listed in Appendix A, which should be accessible online. For this article, we selected the experiments that best demonstrated our findings.

The inset in Fig.\ref{fig:fig1} represents the TPD of 1ML NH$_{3}$ on a gold substrate maintained at 10K and is used to calibrate all the experiments. In the calibration experiment, to determine the dose for 1ML, we performed a series of TPDs with various doses of NH$_{3}$ on a gold substrate held at 10K and a flux of 0.1 sccm in the injection chamber. We found that 1ML of NH$_{3}$ is deposited after $\sim$9min of deposition. The ramp for all the experiments is 0.2K/s. The details of the calibration can be found in Appendix B.

Fig. \ref{fig:fig1} shows the TPD curves of the NH$_{3}$-H$_{2}$O co-deposition experiments. In the calibration experiment, we observe a sudden rise in the NH$_{3}$ desorption around 80K which falls off slowly once peak desorption has been completed. Contrarily, during the co-deposition with water, we observe a significant decrease of the NH$_{3}$ desorption rate, and peak desorption shifted to higher temperatures. We find that about 6$\%$ w.r.t to H$_{2}$O of the deposited NH$_{3}$ is trapped by water and then released during the phase change of water from amorphous to crystalline. This trapped fraction is estimated by calculating the area under the TPD profile of the desired species (in this case NH$_{3}$). In reality, water deposited at such low temperature is fairly porous and, hence, has a larger surface area wherein the NH$_{3}$ molecules can lodge themselves. As the temperature is raised, water undergoes a phase change from amorphous to crystalline which can be observed as the plateau/shoulder at around 155K in the desorption curve of water (Fig. 1, see also \cite{speedy1996evaporation}), as well as a change in the IR spectra (not shown here). During this phase change, water molecules begin to re-arrange to form a crystalline structure during which it pushes out any NH$_{3}$ in its bulk. Furthermore, the higher the H$_{2}$O/NH$_{3}$ ratio, the more the desorption of ammonia is delayed - as expected. Indeed, when the NH$_{3}$ concentration is very high, there are more NH$_{3}$-NH$_{3}$ interactions which can substitute for NH$_{3}$-H$_{2}$O interactions.

\begin{center}
\setlength{\tabcolsep}{10pt}
\renewcommand{\arraystretch}{1.0}
\begin{table*}[h]
\centering
\caption{List of co-deposition experiments.} \label{table:Table1}
\begin{tabular}{cccc}
\hline
\textbf{No.} & \textbf{Experiment} & \textbf{Ratio} & \textbf{\begin{tabular}[c]{@{}c@{}}Quantity \\ deposited (ML)\end{tabular}} \\ [0.2in] \hline 
&&& \\
1 & \{NH$_{3}$ + H$_{2}$O\} & 1:1 &  \begin{tabular}[c]{@{}c@{}}˜10 ML of each \end{tabular}  \\ [0.1in]
2 & \{NH$_{3}$ + H$_{2}$O\} & 1:4 & \begin{tabular}[c]{@{}c@{}}5.8 (NH$_{3}$), \\ 23.8(H$_{2}$O)\end{tabular}  \\ [0.2in] \hline
&&& \\
3 & \{NH$_{3}$ + $^{13}$CO\} & 1:1 & \begin{tabular}[c]{@{}c@{}}11.3(NH$_{3}$),\\ 12.3($^{13}$CO)\end{tabular}  \\ [0.2in]
4 & \{NH$_{3}$ + $^{13}$CO\} & 1:6 & \begin{tabular}[c]{@{}c@{}}1(NH$_{3}$), \\ 6.5($^{13}$CO)\end{tabular} \\[0.2in]
5 & \{NH$_{3}$ + $^{13}$CO + H$_{2}$O\} & 1:1:1 & \begin{tabular}[c]{@{}c@{}}8.1(NH$_{3}$), 9.2($^{13}$CO),\\  9.2(H$_{2}$O)\end{tabular} \\[0.2in]
6 & \{NH$_{3}$ + $^{13}$CO + H$_{2}$O\} & 1:2:9 & \begin{tabular}[c]{@{}c@{}}1.6(NH$_{3}$), 3.6($^{13}$CO),\\  8.9(H$_{2}$O)\end{tabular}  \\ [0.2in] \hline
&&& \\
7 & \{NH$_{3}$ + CO$_{2}$\} & 1:1 & \begin{tabular}[c]{@{}c@{}}14.7(NH$_{3}$),\\ 14.7(CO$_{2}$)\end{tabular}  \\ [0.2in]
8 & \{NH$_{3}$ + CO$_{2}$\} & 1:7 & \begin{tabular}[c]{@{}c@{}}0.7(NH$_{3}$),\\ 5.4(CO$_{2}$)\end{tabular}  \\ [0.2in]
9 & \{NH$_{3}$ + CO$_{2}$ + H$_{2}$O\} & 1:1:1 & \begin{tabular}[c]{@{}c@{}}9(NH$_{3}$), 9.3(CO$_{2}$), \\ 8.9(H$_{2}$O)\end{tabular} \\ [0.2in]
10 & \{NH$_{3}$ + CO$_{2}$ + H$_{2}$O\} & 1:4:5 & \begin{tabular}[c]{@{}c@{}}1.5(NH$_{3}$), 4(CO$_{2}$), \\ 5.5(H$_{2}$O)\end{tabular}  \\ [0.2in] \hline
\end{tabular}
\caption*{\small \textbf{Note:} All experiments are performed on a gold substrate held at 10K. The ratios have been rounded off to the nearest whole number for visual convenience while the no. of mono-layers deposited has been mentioned to one decimal point of accuracy. The ramp during the TPD is 0.2K/s. Additional experiments have been carried out but not included in the main article for brevity and can be found in Appendix \ref{AppendixA}}
\end{table*}
\end{center}

\begin{figure*}[h!] 
\begin{center}
\includegraphics[width=\textwidth]{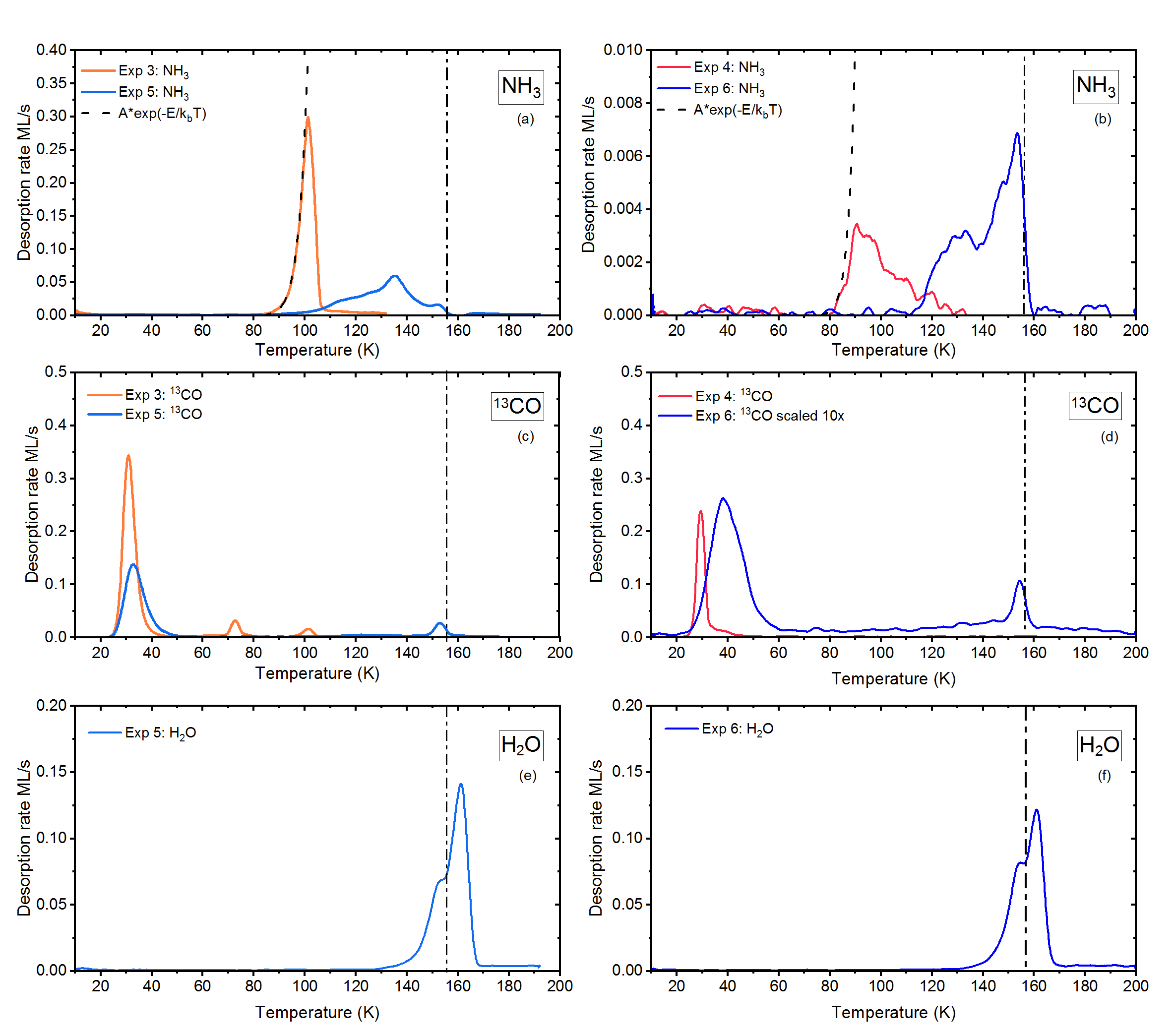} 
\vspace*{-6mm} 
\caption{ \small \textbf{NH$_{3}$-$^{13}$CO (and H$_{2}$O) co-deposition experiments.} The mass channels used for H$_{2}$O, NH$_{3}$ and $^{13}$CO are 18 amu, 17 amu, and 29 amu, respectively also accounting for fragments of mass = 17 amu (for H$_{2}$O) and 16 amu (for NH$_{3}$). All experiments have been performed on a gold substrate. The TPDs have a ramp of 0.2K/s. Lines of the same colour belong to the same set of experiments. The dash-dot vertical line at 155K corresponds to the temperature of the phase change of water from ASW to crystalline.}
\label{fig:fig2}
\end{center}
\end{figure*}

During the TPD of the co-deposition experiments of   NH$_{3}$-$^{13}$CO mixture (Fig. \ref{fig:fig2}), NH$_{3}$ desorbs independently (top row Fig. \ref{fig:fig2}a and \ref{fig:fig2}b) of $^{13}$CO, irrespective of the ratio between the two. All of the NH$_{3}$ is desorbed between 80K-132K. The desorption is in good agreement with a fit to the desorption rate equation \textit{A$\times$exp(-E/k$_{b}$T)}, where E/k$_{b}$ is the binding energy (in Kelvins), T is the temperature (in Kelvins) and A is the pre-exponential factor (s$^{-1}$). Since these experiments are in the multilayer regime (zero order kinetics), we chose E/k$_{b}$ = 2965K and the A = 2.1 $\times$10$^{12}$s$^{-1}$ following \cite{martin2014thermal} and assuming that 1ML = 10$^{15}$ molecules cm$^{-2}$. The offset from the fit seen in \ref{fig:fig2}b is due to the low quantities of NH$_{3}$ used for the experiments. Higher quantities would ensure a good fit as in the previous case. 

When present with NH$_{3}$ in equal quantities, 85.8$\%$ of the deposited $^{13}$CO is desorbed between 20K-50K (Fig \ref{fig:fig2}c middle row orange curve). At the same time, $^{13}$CO exhibits a volcano effect due to NH$_{3}$. A volcano effect is the sudden desorption of a volatile species trapped under a less volatile species like H$_{2}$O when the latter begins to desorb (\cite{smith1997molecular}, \cite{viti2004evaporation}, \cite{collings2003carbon}). This volcano effect can be observed as the peak between 68K-80K and accounts for $\sim$ 9$\%$ of the total desorbed $^{13}$CO. On the contrary, this trapping is not observed when NH$_{3}$ is present in trace quantities, as expected (Fig \ref{fig:fig2}d). The $^{13}$CO peak between 90K-110K is the CO desorbing with NH$_{3}$. This is probably the CO adsorbed on the substrate during the initial moments of the co-deposition and thereby buried under the later layers of NH$_{3}$ and $^{13}$CO. The right panels of Fig. \ref{fig:fig2} contain the experiments in proportions that are more astronomically relevant. Hence, here, NH$_{3}$ is deposited in trace amounts (1 ML or sub-monolayer quantities) as compared to $^{13}$CO and H$_{2}$O. Once again, for NH$_{3}$, we observe a similar desorption trend as in the case of the experiments presented in Fig. \ref{fig:fig1}.

In the presence of H$_{2}$O, the trend in the desorption of NH$_{3}$ and $^{13}$CO is different as a significant delay in the desorption of NH$_{3}$ is observed. When present in quantities roughly equal to water, the NH$_{3}$ desorption rate is slower and the desorption is delayed compared to its desorption in the absence of water. In the multi-layer (Fig. \ref{fig:fig2}a) and the sub-monolayer (Fig. \ref{fig:fig2}b) scenario, the NH$_{3}$ desorption is shifted to higher temperatures. In the former, NH$_{3}$ desorption is delayed possibly because it needs to diffuse through the bulk of the ice before desorption can take place. In the latter, at low concentrations there is a higher probability for greater NH$_{3}$-H$_{2}$O interaction and lower NH$_{3}$-NH$_{3}$ interaction via hydrogen bonds resulting in water holding on to NH$_{3}$ for a little longer than when NH$_{3}$ is present in larger quantities. It is worth noting in both cases that an NH$_{3}$ volcano peak is observed due to the crystallisation of water. Roughly 5$\%$ of NH$_{3}$ w.r.t water is trapped by water when co-deposited in equal quantities and around 8.5$\%$ is trapped by water in the more astronomically relevant scenario (Expt 6 in Table. \ref{table:Table1}) where it is released, as observed in previous experiments, during the change of phase of H$_{2}$O from amorphous to crystalline. 

\begin{figure*}[h!] 
\centering
\includegraphics[width=18cm]{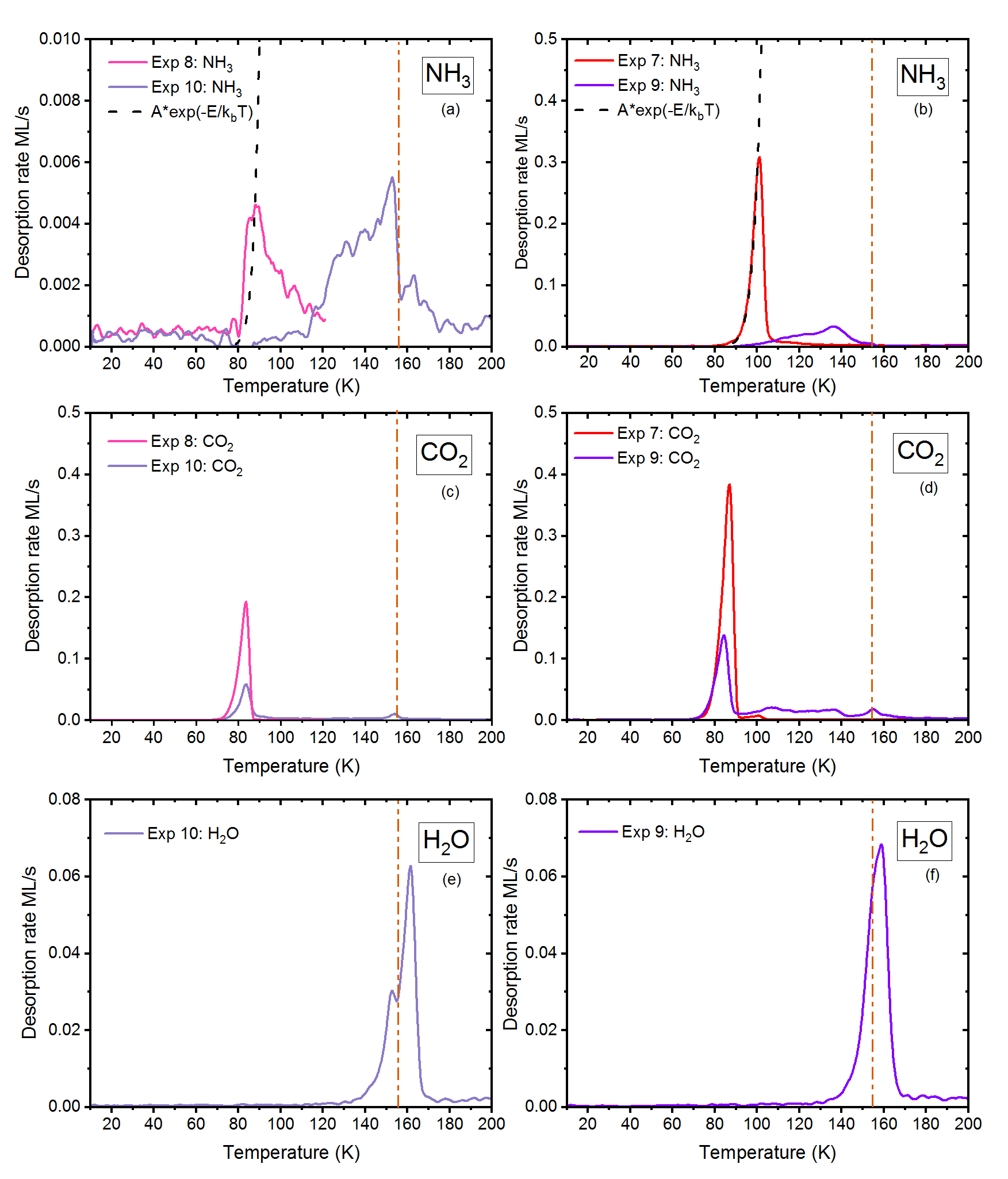} 
\caption{\small Same as Fig \ref{fig:fig2} but using CO$_{2}$ instead of $^{13}$CO. The mass channel used for CO$_{2}$ is 44 amu. }
\label{fig:fig3}%
\end{figure*}

The co-deposition experiments of NH$_{3}$ with CO$_{2}$ also exhibit a similar behaviour as for CO ones. NH$_{3}$ desorption appears to be unaffected by the presence of CO$_{2}$ (Fig.\ref{fig:fig3}a and \ref{fig:fig3}b). NH$_{3}$ desorbs in the same temperature range as previously observed for experiments with $^{13}$CO. Its desorption fits well the curve of desorption rate equation mentioned earlier with the values taken from \cite{martin2014thermal}. When the two are in equal quantities (Fig. \ref{fig:fig3}b), the bulk of CO$_{2}$ desorbs between 70K-92K. 2$\%$ of the deposited CO$_{2}$ desorbs during the desorption of NH$_{3}$. This could be due to mechanical trapping by NH$_{3}$, and hence the former could desorb only after all the NH$_{3}$ desorbed. The shift in peak desorption temperature of NH$_{3}$ towards higher values observed in Fig. \ref{fig:fig3}b as compared to Fig. \ref{fig:fig3}a is due to the fact that in the former we have multiple layers of NH$_{3}$. Additionally, in the sub-ML experiments (Fig. \ref{fig:fig3}a, c and e), we use very low quantities of NH$_{3}$ and therefore, the interaction between NH$_{3}$ and the gold in the substrate becomes significant. Once again, the presence of H$_{2}$O seems to significantly alter the desorption of both NH$_{3}$ (Fig. \ref{fig:fig3}a and b) and CO$_{2}$ (Fig. \ref{fig:fig3}c and d). NH$_{3}$ desorbs in a similar manner to the previous experiments with $^{13}$CO. Its desorption is not only delayed but the rate is also slower. Roughly 9$\%$ of it is trapped and then released during the phase change of H$_{2}$O. CO$_{2}$ desorption, on the other hand, doesn't appear delayed. However, roughly 2$\%$ of it is trapped by water and then released along with NH$_{3}$ during the phase change of H$_{2}$O. In previous experimental works, \cite{Bossa2008} and  \cite{noble2014kinetics} have shown that carbamic acid (NH$_2$COOH) can be formed as early as 80K by thermal reaction of CO$_2$ and NH$_3$. We do not observe this reaction in our experiments - which are performed in much thinner layers - probably because this reaction only occurs if the reactants are solvated. Similarly, \cite{potapov2019evidence} reported the formation of NH$_{4}^{+}$NH$_{2}$COO$^{-}$, which we do not observe in our experiments.

\subsection{Desorption of NH$_{3}$ from different types of ices }

A second set of experiments is performed to understand the desorption dynamics and determine the distribution of binding energies. The method used has been discussed in more detail in \cite{de1990thermal}, \cite{he2011interaction} and \cite{amiau2006}. Subsequent depositions of roughly 1 ML of NH$_{3}$ are made on each type of ice substrate followed by a TPD. Two kinds of ice substrates are used for this purpose: crystalline ice and compact amorphous solid water ice. The crystalline ice substrate is formed by depositing water onto the deposition surface held at 110K followed by flash heating up to 150K. To create a compact amorphous water ice substrate, water is deposited onto the gold substrate at 110K and an incident angle of $\theta$ = 0$\degree$ with respect to the normal of the gold substrate. A normal angle of incidence leads to a denser ice substrate as shown by \cite{10.1063/1.1350581,kimmel2001control}. Several studies have shown a key role between the deposition temperature and its effect on the ice substrate. \cite{scott2006effect} found that water deposited at $\leq$ 110 K is amorphous but already begins to pre-crystallise at $\geq$ 120 K since it is thermodynamically favoured.  \cite{he2019effective} and \cite{bossa2012thermal} observed a significant reduction in the porosity of the substrate beyond 100 K. Both these studies are performed on samples that are $\sim$ 200 MLs and $\sim$ 3000MLs, respectively, while the present study focuses on thin ices ($\sim$ tens of MLs). Hence is safe to assume that the amorphous substrate formed in our case is, indeed, compact. \\

To prepare the raw data for analysis, an initial adjacent averaging smoothing process employing 35 data points is applied using Origin software. This initial smoothing is carried out to reduce background noise which becomes significant at low dosages and high temperatures. Subsequently, the smoothed data is fed into a custom software developed at LERMA, which fits a set of 19 independent TPD curves distributed evenly over a range of 19 binding energies spanning from 3030 K to 5730 K. For a more comprehensive explanation of this method, please refer to Chaabouni et al. (2018). In our analysis, we used a pre-exponential factor, A, with a value of 1.94 x 10$^{15}$ s$^{-1}$, as per the findings of \cite{minissale2022thermal}. The results are presented in the form of blue curves, representing the mass spectrometer data, and orange curves, representing the fit to the experiments conducted using software developed within our laboratory.\\

\begin{figure*}[h!]
    \centering
     \begin{subfigure}[b]{0.8\textwidth}
         \centering
         \includegraphics[width=\textwidth]{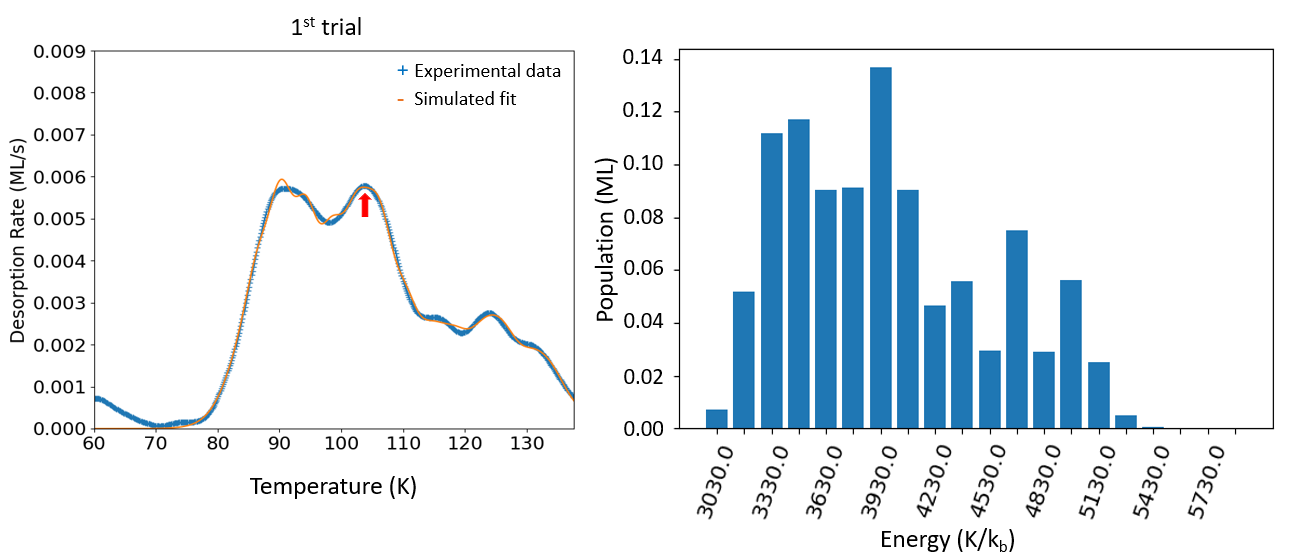}
         \caption{}
         \label{fig:4a}
     \end{subfigure}
    \begin{subfigure}[b]{0.8\textwidth}
         \centering
         \includegraphics[width=\textwidth]{NH3_on_CI_trial_1_TPD_fit_and_BE2.png}
         \caption{}
         \label{fig:4b}
     \end{subfigure}
    \begin{subfigure}[b]{0.8\textwidth}
         \centering
         \includegraphics[width=\textwidth]{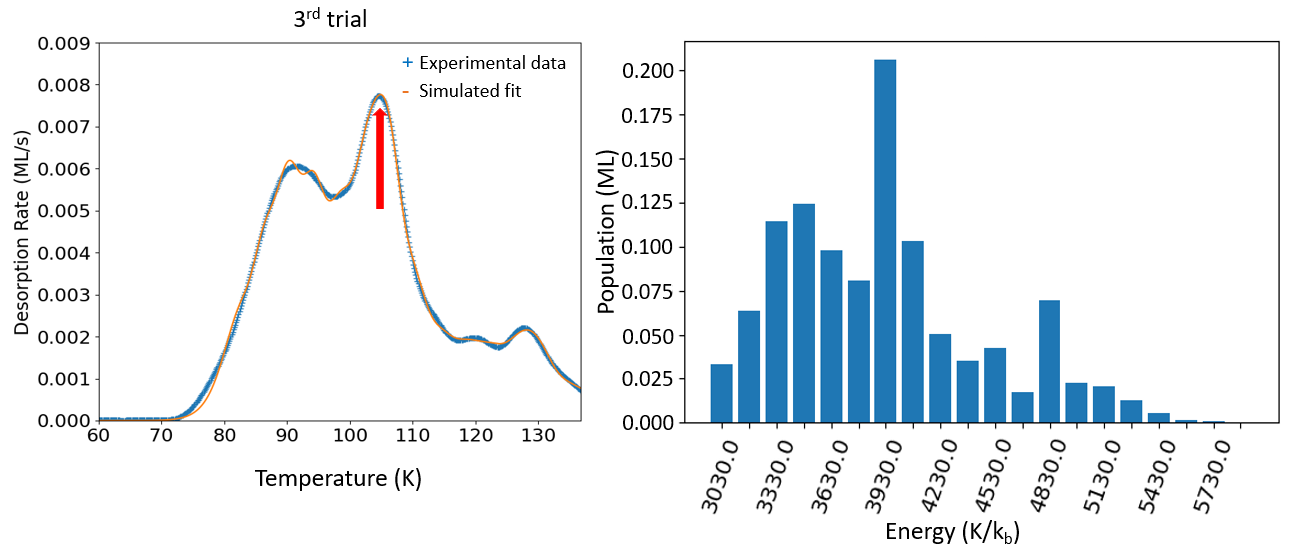}
         \caption{}
         \label{fig:4c}
     \end{subfigure}
    \caption{\small Binding energy fits of the TPDs \textit{(left column)} and the corresponding binding energy distribution histograms \textit{(right column)} of three separate, subsequent depositions of 1 ML of NH$_{3}$ on the surface of crystalline ice. Each deposition is followed by a TPD to remove the ammonia deposited on the ice substrate before the consequent deposition is made. The preference of ammonia to bind to water instead of itself is seen as the progressive increase in peak height (indicated by the red arrow) with each trial. }
    \label{fig:fig4}
\end{figure*}

NH$_{3}$ on CI (Fig. \ref{fig:fig4}) desorbs in a manner similar to NH$_{3}$ on the gold substrate as in the calibration experiments. There is a sharp increase in the desorption rate at 75K. Most of the desorption occurs between 78K and 140K. On CI, we observe two peaks. The first one (between 76K - 98K) is the multi-layer desorption of NH$_{3}$ due to its interaction with itself rather than with the surface and does not necessarily need full layers of NH$_{3}$ to be present to appear. The second one is between 98K - 112K and we observe an increase in the height of this peak with each deposition. This implies that the surface evolves with each cycle of dosing the crystalline surface with NH$_{3}$. In other words, NH$_{3}$ is able to amorphise the surface structure of the ice by introducing defects onto its surface. This modification seems, however, to be a surface phenomenon and alters only the top layers of the surface. Even when the quantity of NH$_{3}$ is increased, NH$_{3}$ prefers to bind to H$_{2}$O than with itself as is evident by this observed increase in peak height in Fig. \ref{fig:fig4}. Nevertheless, this modification doesn't affect the desorption of either H$_{2}$O or NH$_{3}$ as the latter is eventually pushed out of the surface of CI during the desorption of water. This, in turn, renders it harder to calculate one single value of binding energy for NH$_{3}$ on CI. \\

\cite{he2016binding} conducted a similar investigation in which they deposited NH$_{3}$ onto a CI substrate to determine the binding energy of NH$_{3}$ desorption. Our study exhibits considerable resemblance with their research, particularly with respect to their TPD curve obtained for a deposition of 2ML. Noticeably, a major proportion of the desorption events in both studies occur within the temperature range of 80 K to 145 K. However, a discrepancy arises in the temperature at which the multilayer desorption peak is observed. He et al. reported this peak occurring at a slightly elevated temperature, approximately 100 K. This difference could potentially be attributed to their use of a higher heating ramp rate, estimated to be around 0.5 K/s, during their TPD experiments. Another distinguishing feature between the two studies pertains to the temperature of complete desorption of the ices. In our investigation, it was observed that the ices are nearly completely desorbed by 140 K, whereas He et al. noted the presence of desorption signals persisting beyond 140 K. Unfortunately, a more comprehensive comparison is hindered by the lack of specified units for the desorption rate in the pertinent figure within their study. But the difference in heating speed is certainly the main difference, since the slower the heating, the earlier the desorption, for an equivalent quantity.
In terms of the determination of binding energies (BE), He et al., adopting the direct inversion method, reported BE values falling within the range of approximately 2900 K to 4100 K. Notably, these BE values are lower than the values obtained in our study. This variance in BE values may be attributed to the use of a lower pre-exponential factor, specifically 10$^{12}$s$^{-1}$, in their methodology as compared to the one employed in our study. Using the conversion formula proposed in \cite{chaabouni2018thermal}, reported in the review of  \cite{minissale2022thermal}, we find a good agreement.\\

\begin{figure*}[h!]
    \centering
    \includegraphics[width=6.8in]{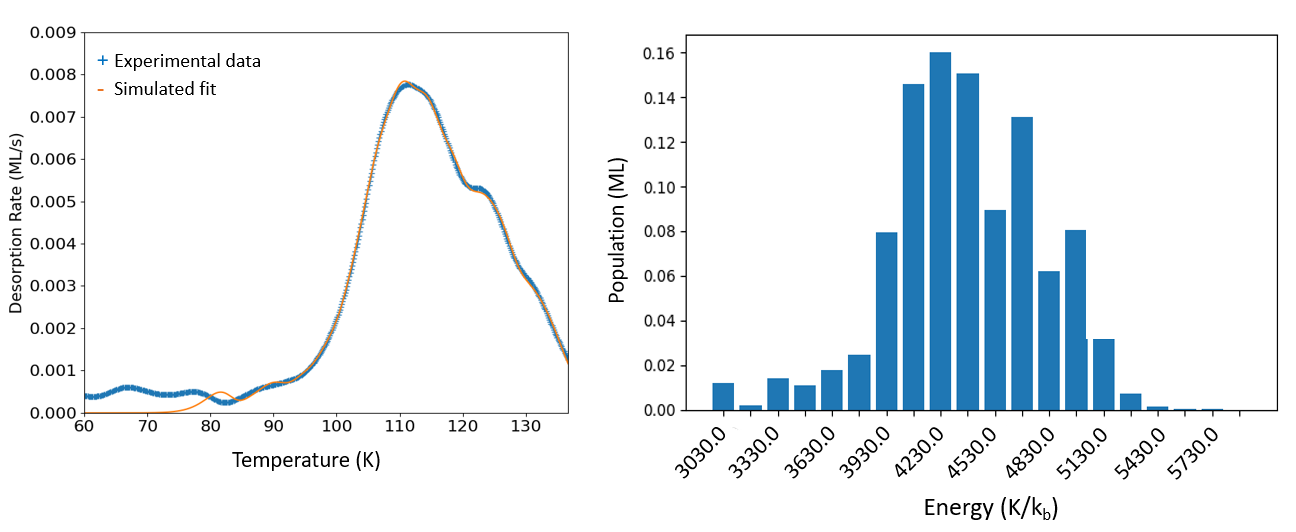}
    \caption{\small Binding energy fit \textit{(left panel)} for 1ML deposition of NH$_{3}$ on c-ASW ice and the corresponding binding energy histogram \textit{(right panel)}.}
    \label{fig:5}
\end{figure*}

In contrast, the desorption of NH$_{3}$ on c-ASW (Fig. \ref{fig:5}) occurs at higher temperatures. This delay could be associated with the structure of c-ASW i.e., due to the presence of concavities at the molecular level on its surface where NH$_{3}$ can lodge itself and be surrounded by many water molecules and hence form more hydrogen bonds than in the case of CI. We observe a single broad peak, which is in contrast to what we observe with CI. A bulk of this desorption occurs between 95 K and 140 K. On crystalline ice, we obtained the binding energy between two mono-layers of ammonia in the range of 3180K- 3630K and between ammonia and the surface of the water substrate in the range of 3780K – 4080K. On c-ASW, we obtained binding energy values between 3630K and 5280K. Our BE values are in very good agreement with the values obtained on ASW ice via theoretical calculations by \cite{tinacci2022theoretical} who also used the same value of the pre-exponential factor. We also note that in experiments as well as calculations, binding energy distributions are important, and have a very good match with \cite{Germain2022}. A more detailed discussion between the comparison of experiments and calculations derived values can be found in \citep{Ferrero2022}.

\subsection{Temperature Programmed-During Exposure Desorption (TP-DED) experiments}

\begin{figure*}
    \centering
    \includegraphics[width=16cm]{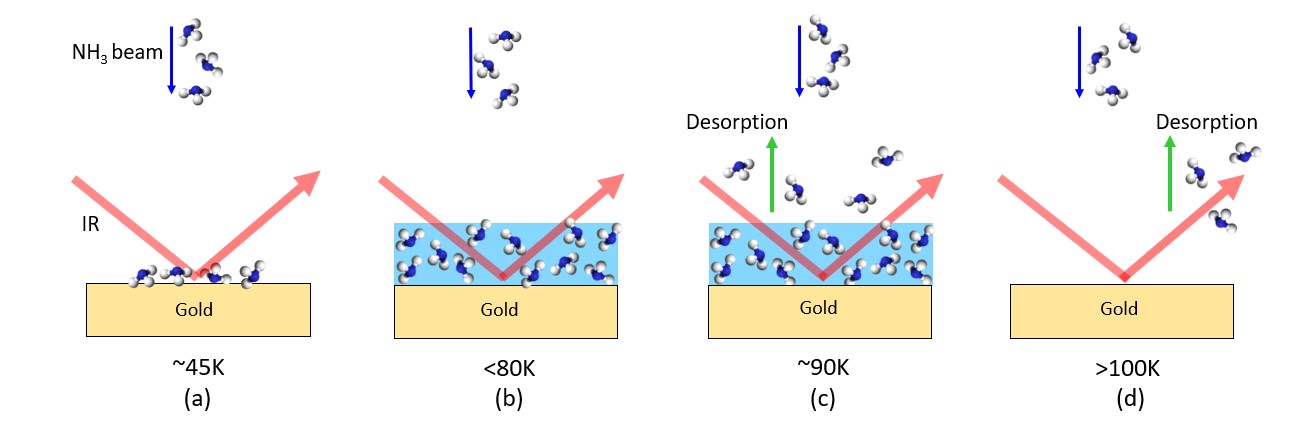}
    \caption{\small Schematic explaining the sequence of events in the TP-DED experiment. Here, we use the example of an NH$_{3}$ beam but the same procedure follows for all three TP-DED experiments performed in this work. The temperatures in each sub-figure in the schematic are merely indicative due to the flux dependence of the adsorbate and not the unique temperature at which the events take place. }
    \label{fig:figure 6}
\end{figure*}

\begin{figure*}[]
    \centering
     \begin{subfigure}[b]{\columnwidth}
         \centering
         \includegraphics[width=9cm]{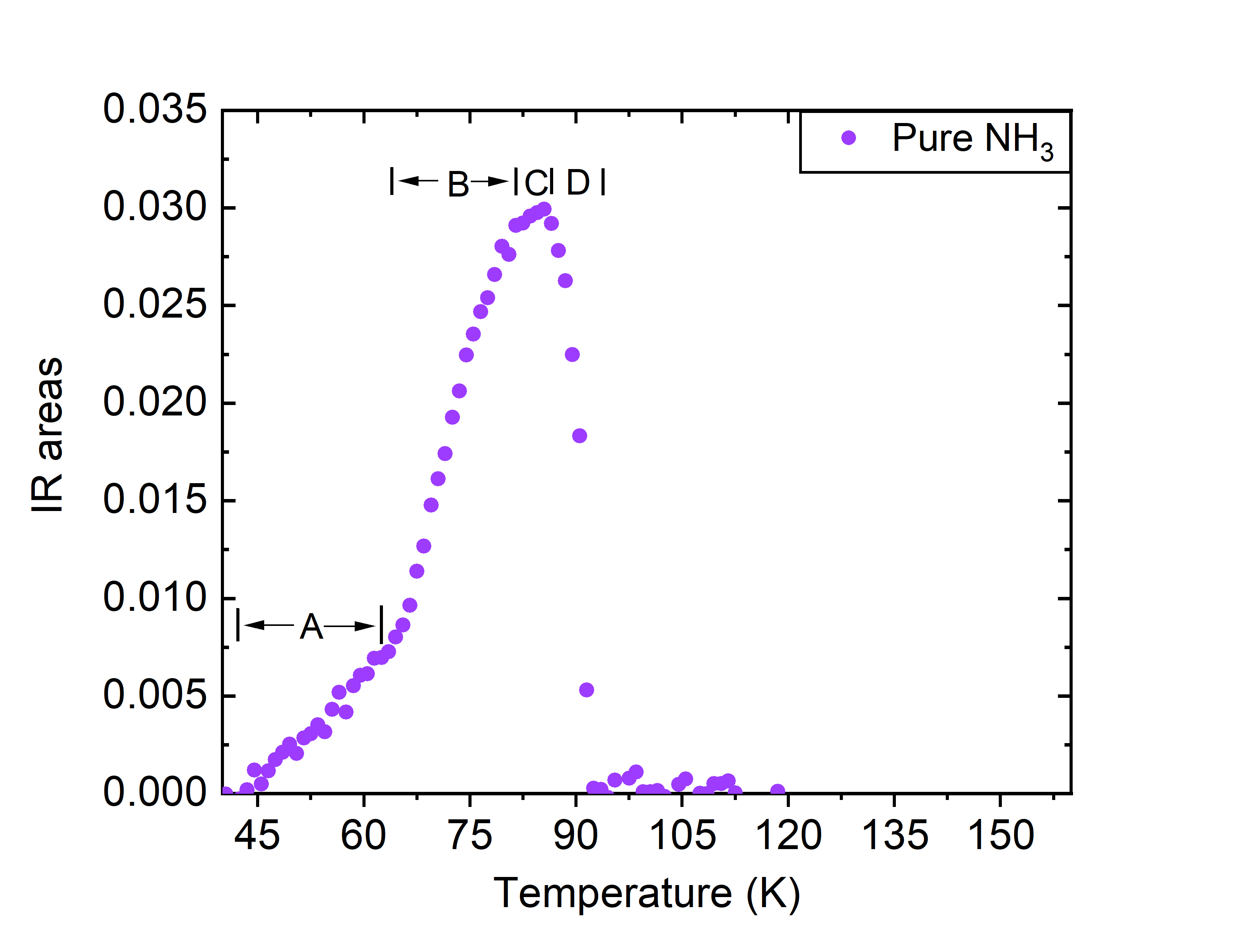}
         \caption{}
         \label{fig:7a}
     \end{subfigure}
    \begin{subfigure}[b]{0.9\columnwidth}
         \centering
         \includegraphics[width=9cm]{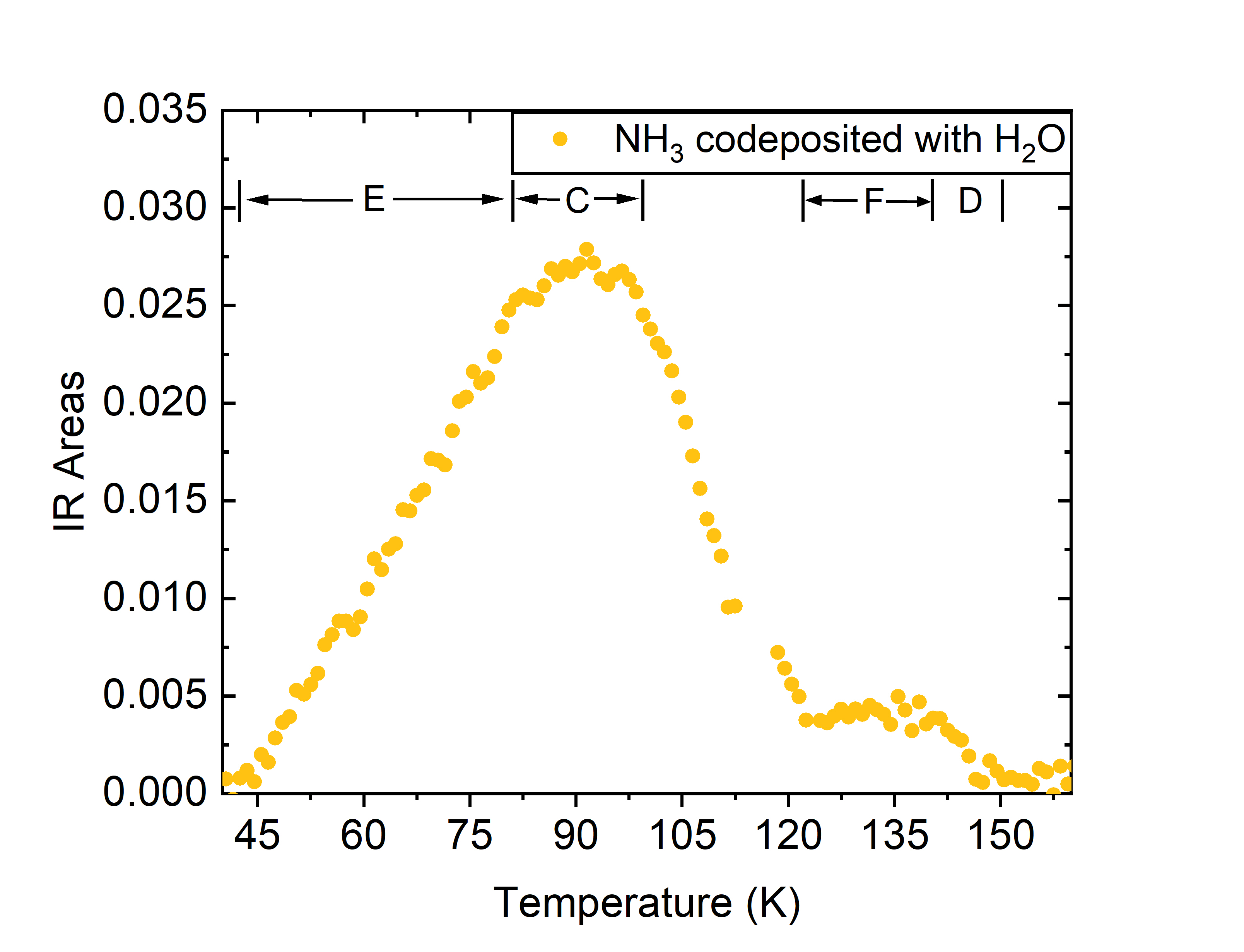}
         \caption{}
         \label{fig:7b}
     \end{subfigure}
    \begin{subfigure}[b]{\columnwidth}
         \centering
         \includegraphics[width=10cm]{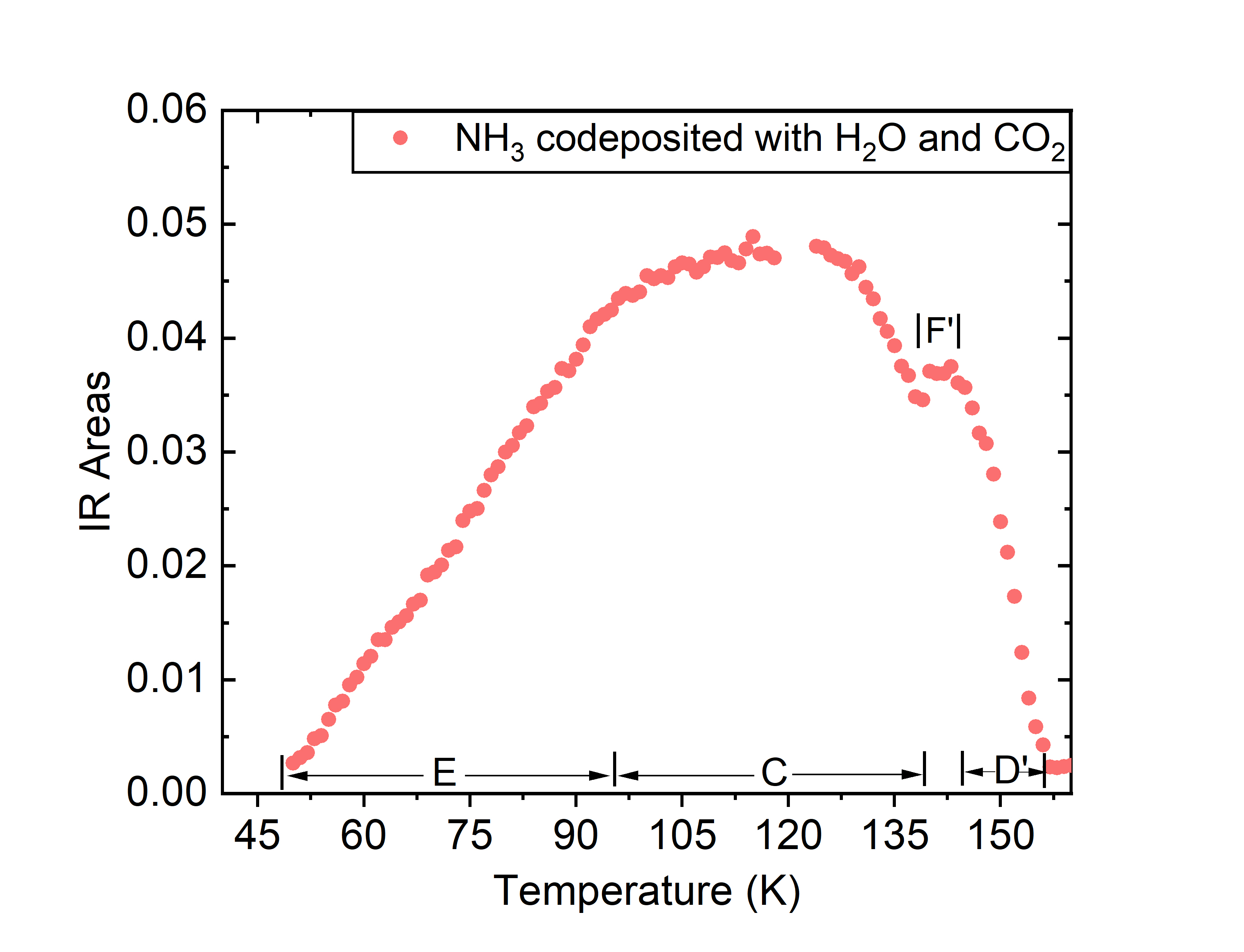}
         \caption{}
         \label{fig:7c}
     \end{subfigure}
    \caption{\small Figure depicting the quantity of NH$_{3}$ deposited, with respect to temperature (K), during the TP-DED experiments measured using a FT-RAIRS. For details on the zones A, B, C, D, D', E, F AND F', refer to main text Section 3.3}
    \label{fig:7}
\end{figure*}

Fig. \ref{fig:figure 6} shows the basic schematics of a TP-DED experiment using the example of pure NH$_{3}$ deposition. For this set of experiments, the desired species is deposited onto the gold substrate Fig. \ref{fig:figure 6}a. simultaneously while the latter is heated/cooled at a constant ramp of 0.5K/min. During the heating process, as the deposition progresses, the species continues to accumulate/adsorb on the substrate \ref{fig:figure 6}b. until the desorption temperature range of the former is reached. At this point, the species begins to slowly desorb from the substrate, and any incoming molecule of the species desorbs immediately \ref{fig:figure 6}c. until no more incoming molecules can adsorb due to the temperature of the surface being higher than the desorption temperature \ref{fig:figure 6}d. The reverse is true during the cooling process. The experiment is monitored using the FT-RAIRS during the deposition and is usually followed by a TPD at the end. The details of each experiment are given in Table \ref{label:Table2}. Here, this rather rarely used type of experiment aim at mimicking snow line regions where both accretion and depletion on grains (i.e. sticking and desorption), are taking place simultaneously.  A snow line is defined in a region with a thermal gradient that applies on the grains (\emph{i.e.,} hotter towards the star and cooler in the outer regions) such that there exists a frontier zone (the snow line) where the accretion compensates for the desorption. Towards the colder side, the accretion dominates and the molecules condense onto the grains whereas on the hotter side the concerned species are in the gas phase. Our goal is to investigate the snow line of NH$_{3}$. The results of the experiments can be found in Fig. \ref{fig:7}.\\

\begin{table}[]
\centering
\caption{\small List of TP-DED experiments}\label{label:Table2}
\begin{tabular}{cccl}
\cline{1-3}
Expt No. & Adsorbate       & \begin{tabular}[c]{@{}c@{}}Temperature\\ Range (K)\end{tabular}  \\ \cline{1-3}
1        & NH$_{3}$             & \begin{tabular}[c]{@{}c@{}}\\40 - 105\\ 105 - 94\end{tabular}  \\
&&& \\
2        & NH$_{3}$ + H$_{2}$O       & \begin{tabular}[c]{@{}c@{}}40 - 180\\ 180 - 60\end{tabular} \\
&&& \\
3        & NH$_{3}$ + H$_{2}$O + CO$_{2}$ & 40 - 180                                          \\[0.2in] \cline{1-3}

\end{tabular}
\end{table}

Our study plots the area under the IR peak at 3383 cm$^{-1}$ of NH$_{3}$ against temperature. The areas are calculated using the vibrational spectroscopy software 'OPUS' by the company BRUKER. We apply a baseline correction where necessary. It is important to mention that in experiments 2 and 3, due to heavy baseline distortion and the fact that the IR peaks of NH$_{3}$ were superimposed with that of H$_{2}$O, we were obliged to make gross assumptions to obtain the IR area of NH$_{3}$. The area we report is, hence, the area of NH$_{3}$ subtracted from the combined areas of NH$_{3}$ and H$_{2}$O.

All three experiments (Fig. \ref{fig:7a}, \ref{fig:7b} and \ref{fig:7c}) show distinct zones during the TP-DED. In the first experiment Fig. \ref{fig:7a}, Zone A corresponds to the accretion of NH$_{3}$ in the amorphous form. Given the time to form 1ML on the gold substrate, the signal before 60K corresponds to NH$_{3}$ on the substrate. Beyond this temperature, the signal observed is of NH$_{3}$ on this NH$_{3}$ layer. Since the surface is continuously exposed to gaseous NH$_{3}$ and the desorption is negligible for these temperatures, ammonia accumulates on the surface, and we observe a linear increase in the IR absorbance signal. The same trend would be observed for a constant surface temperature desorption.
As the temperature rises, NH$_{3}$ begins to change its structure from amorphous to crystalline, denoted as zone B and can be identified by a change in the slope. This is not due to a change in the flux of accretion - which is stable during every experiment - but is due to an increase in the absorption band strength after the phase transition. This phenomenon is also observed in \cite{cazaux2021photoprocessing}, where a similar change is found during the heating of H$_{2}$S. \\

Around 81.5K, we notice a slowing down of the rise, followed by a plateau (Zone C). Even though the desorption rate of NH$_{3}$ increases exponentially, it is in competition with the accretion rate. As a result, a part of the molecules adsorbed is now desorbing, and thus the net accumulation slows down to almost null at the snow line - which is the turning point of the curve at around 88K. Above 90K, we notice a sudden drop in the NH$_{3}$ desorption (zone D). Here, despite the accretion, desorption still dominates. This occurs since there is no more measurable NH$_{3}$ on the surface as the adsorbed ammonia is desorbed quickly after adsorption, which is even faster in a laboratory measurement time frame.\\

We then perform a co-deposition of NH$_{3}$ with H$_{2}$O (see Fig.\ref{fig:7b}) under similar conditions. The first striking observation is the absence of the region of phase change of NH$_{3}$ from amorphous to crystalline. This could be due to the aforementioned preference of NH$_{3}$ to form hydrogen bonds with H$_{2}$O than with itself. Hence, zone E could be considered as the region that contains NH$_{3}$ within the H$_{2}$O structure. The turn-off point between zone E and zone C is less pronounced than in the previous case and marks desorption of NH$_{3}$ in interaction with itself and probably not the water ice. After roughly 105K, there is a decrease in the quantity of NH$_{3}$ due to the substrate reaching a higher desorption flux of NH$_{3}$. However, an interesting point to note here is that there still is a significant quantity of NH$_{3}$ left on the surface. This can be verified by the plateau (zone F) which is an indicator of NH$_{3}$ in H$_{2}$O. This is due to the trapping phenomena observed earlier by H$_{2}$O which is further strengthened by the NH$_{3}$-H$_{2}$O interaction. The slow decrease in the slope in zone D is the release of NH$_{3}$ due to the desorption of H$_{2}$O as the latter approaches its crystallisation temperature. \\

In the next set of experiments (Fig.\ref{fig:7c}), we add CO$_{2}$ to this mixture. During the experiment, we perceive a similar trend to that of the previous experiment with H$_{2}$O at the beginning of the desorption process i.e., the sudden change in the curve for NH$_{3}$ is very subtle, if at all present (zone E). The accretion continues until the mixture reaches around 97K. At this temperature, we begin to recognise some similarities noted in the previous two experiments. Firstly, the signal of the solid NH$_3$ is characterised by a plateau (zone C) due to the competition between NH$_{3}$ accretion and desorption. This is followed by a sudden yet short drop in the quantity of NH$_{3}$ owing to the high temperature. Zone F marks the narrow plateau due to the NH$_{3}$-H$_{2}$O-CO$_{2}$ interaction. It is interesting to note that the quantity of NH$_{3}$ (in other words, the IR area) in this zone is significantly higher than in the previous two cases. This could point to an additional interaction of NH$_{3}$ with CO$_{2}$ that retains NH$_{3}$ until a higher temperature. However, the most interesting point to note between all three experiments is the progressive shift in the peak desorption temperature of NH$_{3}$ as we add each new species to pure NH$_{3}$ - especially in the presence of H$_{2}$O, and even more so with the addition of CO$_{2}$. There is still some NH$_{3}$ in the solid phase at unexpectedly high temperatures. In the case of NH$_{3}$/CO$_{2}$/H$_{2}$O mixture, the snow line (the turning point of NH$_{3}$) coincides with the beginning of the desorption of water, almost superimposing their snow lines. This also raises a question about the potential role of CO$_{2}$ in the desorption of NH$_{3}$. These experiments directly demonstrate that NH$_{3}$ lacks a snow line of its own when combined with H$_{2}$O (and potentially CO$_{2}$) which, as a consequence, retains the NH$_{3}$ molecules on the grain to higher temperatures than previously expected and therefore, in a different spatial zone (the same as water).\\

\section{Conclusions}

Nitrogen-based molecules have been found in the densest parts of the interstellar medium - like pre-stellar cores - where they freeze out at densities higher than those found for CO and other C-bearing molecules. Therefore, N-bearing molecules are excellent tracers of these dense regions that are otherwise difficult to observe. In this present work, we choose NH$_{3}$ and study its interaction on grain surfaces - which is still poorly understood - in the presence of other species found on grain surfaces like H$_{2}$O, CO$_{2}$ and $^{13}$CO. We performed various types of experiments with the four species under conditions that mimic those found in pre-stellar cores and protoplanetary disks, paying close attention to the behaviour of NH$_{3}$ in each case. We also calculated the binding energy of NH$_{3}$ on two different types of water substrates- compact-Amorphous Solid Water Ice (c-ASW) and Crystalline Ice (CI). The main findings of our study are the following: 

\begin{enumerate}
    \item During the co-deposition of NH$_{3}$ with H$_{2}$O, we observe a delay in the desorption and lowering of the NH$_{3}$ desorption rate. Additionally, H$_{2}$O traps around 6$\%$ of NH$_{3}$ which is then released during the water phase change from amorphous to crystalline. On the other hand, when NH$_{3}$ is co-deposited with either $^{13}$CO or CO$_{2}$, we observe no such behaviour. However, this behaviour is observed once again when we add H$_{2}$O to the NH$_{3}$-$^{13}$CO or NH$_{3}$-CO$_{2}$ mixture. In the NH$_{3}$-$^{13}$CO-H$_{2}$O and the NH$_{3}$-CO$_{2}$-H$_{2}$O experiments, we observe roughly 5-9$\%$ of trapped NH$_{3}$ w.r.t water which is then released during the phase change of water from amorphous to crystalline.
    \item  We note that NH$_{3}$ has a range of binding energy values instead of a unique value, in agreement with recent theoretical calculations. On CI, we obtained NH$_{3}$-H$_{2}$O binding energy values in the range of 3780K-4080K. NH$_{3}$ is able to amorphise the substrate surface by disrupting the structural order of the surface of the ice via hydrogen bonding with H$_{2}$O. In the case of c-ASW, we obtained binding energies in the range of 3630K-5280K (for a pre-exponential factor set to A= 1.94 10$^{15}$ s$^{-1}$ in both cases). 
    \item During the TP-DED experiments, the crystallisation of NH$_{3}$ is noticeably impacted in the presence of H$_{2}$O and CO$_{2}$. Furthermore, the desorption temperature of NH$_{3}$ increases significantly in their presence and it desorbs over a longer range of temperatures. There is also some trapping of NH$_{3}$ observed during the experiments. This indicates that NH$_{3}$ has no definite snow line and seems to be strongly influenced by these species. The trapping enables it to be stored on the dust grains and, thereby, be available at later times and/or be transported to higher temperatures (e.g., closer to the central protostar or towards the inner disk of more evolved young stellar objects) to form more complex molecules. 
\end{enumerate}
\begin{acknowledgements}
     This work was funded by CY Initiative of Ex-cellence (grant "Investissements d'Avenir" ANR-16-IDEX-0008), Agence Nationale de la recherche (ANR) SIRC project (Grant ANR-SPV202448 2020-2024), by the Programme National "Physique et Chimie du Milieu Interstellaire" (PCMI) of CNRS/INSU with INC/INP co-funded by CEA and CNES, and by the DIM-ACAV+, a funding programme of the Region Ile de France.
\end{acknowledgements}

\bibliographystyle{aa}
\bibliography{BiblioNH3}


\begin{appendix}

\section{ }

\begin{table}[h]
\centering
\caption{\small Additional experiments carried out to study the behaviour of ammonia with H$_{2}$O, $^{13}$CO and CO$_{2}$ but not included in the main text.}\label{AppendixA}
\setlength{\tabcolsep}{3pt}
\renewcommand{\arraystretch}{1.2}
\begin{tabular}{cccc}
\toprule
\hline 
\\
\textbf{No.} & \textbf{Experiment} & \textbf{Ratio} & \textbf{\begin{tabular}[c]{@{}c@{}}Quantity Deposited\\ \\ (ML)\end{tabular}} \\ \hline \\
1            & \{NH$_{3}$ + H$_{2}$O\}       & 1:3            & \begin{tabular}[c]{@{}c@{}}5(NH$_{3}$),\\ 16 (H$_{2}$O)\end{tabular}                    \\
2            & \{NH$_{3}$ + H$_{2}$O\}       & 1:1            & ~5 of each                                                                     \\ \\ \hline \\
3            & \{NH$_{3}$ + CO$_{2}$ + $^{13}$CO\} & 1:3:15         & \begin{tabular}[c]{@{}c@{}}7.4 ($^{13}$CO),\\ 1.7 (CO$_{2}$),\\ 0.5 (NH$_{3}$)\end{tabular}  \\ \\  \hline \\
4            & \{NH$_{3}$ + CO$_{2}$\}       & 1:3            & \begin{tabular}[c]{@{}c@{}}6 (NH$_{3}$),\\ 18 (CO$_{2}$)\end{tabular}                   \\ \\\hline 
\end{tabular}\par
\end{table}

\section{}

\begin{figure}[h!]
    \includegraphics[width = \linewidth]{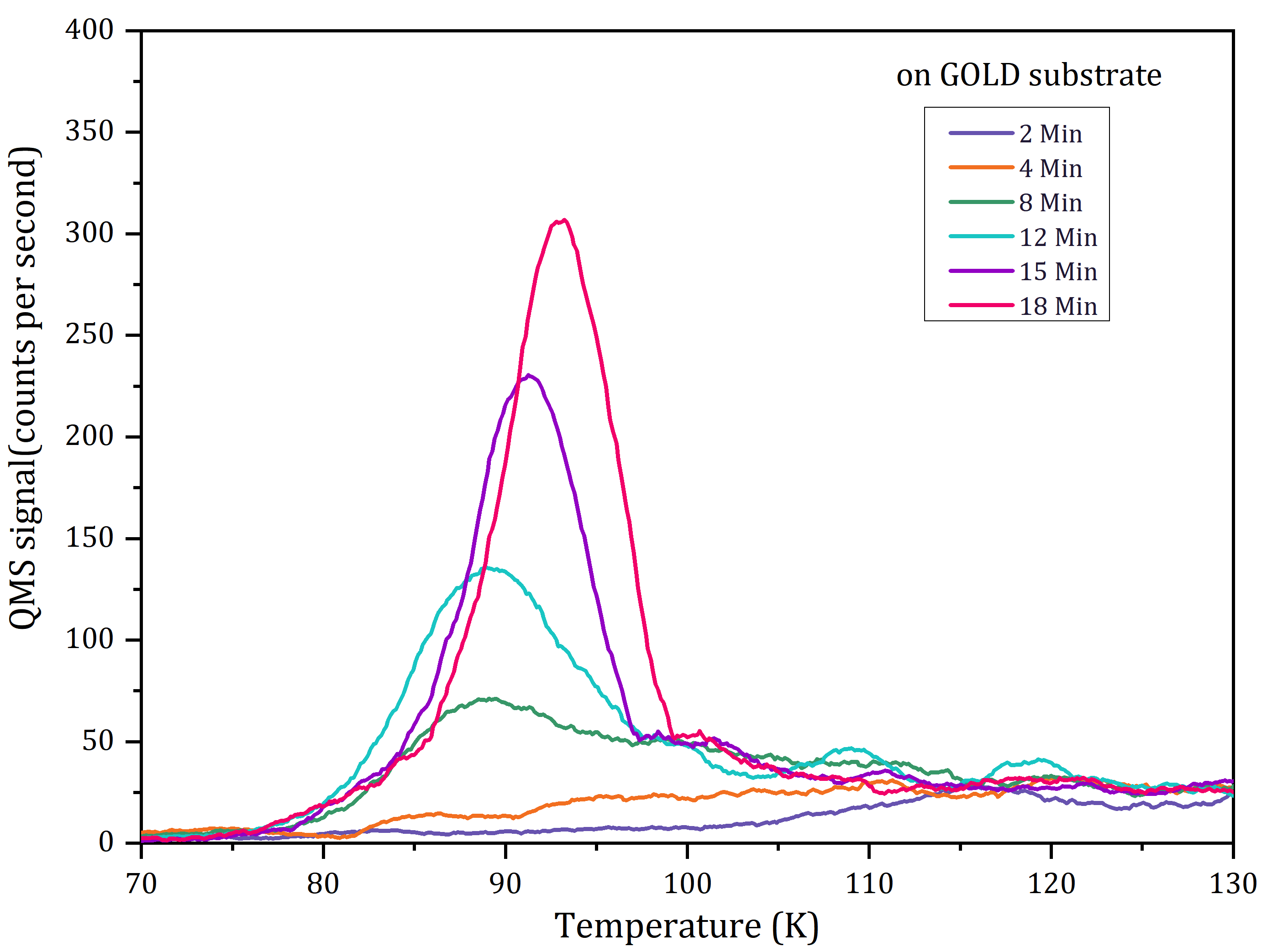}
    \caption{\small Experiments to calibrate for 1ML of NH$_{3}$. TPD curves of NH$_{3}$ deposited on a gold substrate for various dosages, used to calibrate for 1 ML. We use a constant flux of (insert value) and vary the duration of injection to vary the dosage.    }
    \label{fig:calib}
\end{figure}

Fig. \ref{fig:calib} shows the experiments performed to calibrate for 1 ML of NH$_{3}$. Varying doses of NH$_{3}$ are deposited onto the gold substrate at 10K, followed by a TPD. During deposition, adsorption sites with the highest binding energy are occupied first, followed by sites of lower binding energy until all the sites on the substrate are occupied. Hence, a species adsorbed on a site with higher binding energy will desorb at a higher temperature as opposed to desorption at lower temperatures for a site with lower binding energy. Following this principle, at low dosages (< 1ML), peak desorption occurs at higher temperatures. As the dosage increases, this peak shifts to lower temperatures. Once every adsorption site on the substrate is occupied, the peak stops shifting to lower temperatures. At this stage, any other incoming NH$_{3}$ binds to another NH$_{3}$ already adsorbed on the surface, and the desorption then follows zeroth order kinetics. This can be seen as the increase in peak height (12 min curve and above) and the subsequent shifting of the curve towards higher temperatures. Our experiments obtain 1 ML for an exposure time higher than 8 minutes but lower than 12 minutes. Hence, we chose a time of 9 min, similar to the time we obtained for $^{13}$CO and N$_{2}$ under similar beam conditions. \\

\end{appendix}

\newpage

\end{document}